\documentclass[runningheads]{llncs}
\usepackage[T1]{fontenc}
\usepackage{graphicx}
\usepackage{amssymb}
\usepackage{algorithmic}
\usepackage{textcomp}
\usepackage{xcolor}
\usepackage{calc}
\usepackage{tcolorbox}
\usepackage{multirow}
\usepackage{url}
\usepackage{hyperref}
\usepackage{pgfplots}
\pgfplotsset{compat=1.18}
\usepackage{placeins}
\usepackage{fontawesome}
\usepackage{comment}
\usepackage{tabularx}
\usepackage{enumitem}
\usepackage{booktabs}
\usepackage{framed}
\usepackage{fancybox}
\usepackage{makecell}
\usepackage{csquotes}
\usepackage{orcidlink}
\usepackage{multicol}
\usepackage{colortbl}

\begin{document}
\title{Do Users' Explainability Needs in Software Change with Mood?}
\author{Martin Obaidi\orcidlink{0000-0001-9217-3934} \and
Jakob Droste\orcidlink{0000-0001-8746-6329} \and
Hannah Deters\orcidlink{0000-0001-9077-7486} \and
Marc Herrmann\orcidlink{0000-0002-3951-3300} \and
Jil Klünder\orcidlink{0000-0001-7674-2930} \and
Kurt Schneider\orcidlink{0000-0002-7456-8323}}
\authorrunning{Obaidi et al.}
\institute{Leibniz University Hannover, Software Engineering Group, Hannover, Germany \\
\email{\{martin.obaidi, jakob.droste, hannah.deters, marc.herrmann, \\jil.kluender, kurt.schneider\}@inf.uni-hannover.de}}
\maketitle              % typeset the header of the contribution
\begin{abstract}
Context and Motivation: The increasing complexity of modern software systems often challenges users' abilities to interact with them. Taking established quality attributes such as usability and transparency into account can mitigate this problem, but often do not suffice to completely solve it. Recently, explainability has emerged as essential non-functional requirement to help overcome the aforementioned difficulties. 
Question/problem: User preferences regarding the integration of explanations in software differ. Neither too few nor too many explanations are helpful. In this paper, we investigate the influence of a user's subjective mood and objective demographic aspects on explanation needs by means of frequency and type of explanation.
Principal ideas/results: Our results reveal a limited relationship between these factors and explanation needs. Two significant correlations were identified: Emotional reactivity was positively correlated with the need for UI explanations, while a negative correlation was found between age and user interface needs. %No significant correlations were identified for general mood (sentiment) or gender.
Contribution: As we only find very few significant aspects that influence the need for explanations, we conclude that the need for explanations is very subjective and does only partially depend on objective factors. These findings emphasize the necessity for software companies to actively gather user-specific explainability requirements to address diverse and context-dependent user demands. Nevertheless, future research should explore additional personal traits and cross-cultural factors to inform the development of adaptive, user-centered explanation systems. 

\end{abstract}
\section{Introduction}
\label{sec:intro}

Modern software systems are becoming increasingly complex and opaque \cite{Antinyan2020complex,levy2021understanding,Mens2012complex}. Thus, users often face difficulties when using and understanding everyday technologies \cite{Andrade2019usercomplex,Gabbas2021usercomplex}. A widely adopted strategy to address these challenges is to integrate explainability, which aims to make systems more accessible and easier to use~\cite{deters2024UXandExplainability,kohl2019explainability}. However, misplaced or unnecessary explanations can have negative effects, such as increasing users' cognitive load and causing technostress~\cite{chazette2020explainability,deters2024UXandExplainability}. Thus, understanding a user's actual needs for explanations is a cornerstone of explainability engineering~\cite{droste2024explanations,Kim2023whatwhenexplain,kohl2019explainability,unterbusch2023explanation}. %These needs, however, differ for each user within a system and can vary in intensity \cite{chazette2020explainability,droste2023designing}. 

The need for explanations when using software varies widely among individuals: while some users seek detailed insights to completely understand software behavior, others are satisfied with minimal information~\cite{droste2023designing,ramos2021modeling}. Furthermore, it is known that end-users' assessment of usability, a non-functional requirement closely related to explainability, is influenced by their personality~\cite{kortum2018impact}. This suggests that the desire for explanations is influenced by subjective factors such as mood and personal characteristics as well. For instance, a user in a negative mood might experience frustration more easily~\cite{xu2019roleOfUserMood}, potentially leading to a higher need for explanations compared to someone in a neutral or positive mood. Understanding these individual differences is essential for designing personalized user support systems.

Although prior research has explored methods for detecting and categorizing explanation needs within requirements engineering~\cite{chazette2022framework,deters2023ondemand,droste2024explanations}, the role of mood and demographic characteristics in shaping these needs has received little attention. To address this gap, we investigate correlations between mood and demographic factors--such as age or gender-- on the one hand and explanation needs in the context of everyday software use on the other hand. The aim is to determine whether explanation needs systematically vary with users' emotional states or demographic profiles, which could inform the development of user personas based on these attributes.

The insights from this study offer a foundation for software developers and companies to enhance the user experience by tailoring explanations to individual preferences. Specifically, this paper contributes a framework for identifying explanation needs based on mood and demographic factors, providing a basis for heuristics that guide personalized user support. This approach could enable more adaptive and responsive explanations, fostering a user experience that resonates with diverse individual needs and preferences.

The paper is structured as follows: In Section~\ref{sec:background}, we present related work and background details. The development of the study design is described in Section~\ref{sec:research}. Section~\ref{sec:results} summarizes the results, which are discussed in Section~\ref{sec:discussion}, before concluding the paper in Section~\ref{sec:conclusion}.

\section{Background and Related Work}
\label{sec:background}

In this section, we present background information and related work on explainability in software engineering.

\subsection{Explainability}
\label{sec:erklearbarkeit}

Explainability is a non-functional requirement that addresses explanations regarding different aspects of software systems~\cite{chazette2021exploring}. The rise of AI systems led to an increasing demand for explanations regarding the inner workings of these so-called black boxes~\cite{das2020XAISurvey}. However, since everyday software systems are becoming progressively complex independently of AI, explainability is no longer limited to explaining the inner workings of AI systems~\cite{brunotte2023privacy,chazette2022framework,deters2024qualitymodel,obaidi2025appKonwledge,obaidi2025automatingexplanationneedmanagement,droste2024explanations}. 

Droste et al.~\cite{droste2024explanations} conducted an online survey and revealed that there is a pressing need for explanations in several kinds of everyday software systems. They created a taxonomy that divides the expressed needs for explanation into five main categories. The most frequently mentioned category was the need for explanations regarding interactions with the system. This includes how certain operations can be performed or how to navigate to certain views or functionalities. The second category they describe is \textit{system behavior}, including the need for explanations regarding unexpected system behavior or consequences that arise from the user's actions. \textit{Domain knowledge} concerns questions about terminology or system-specific elements. The forth category covers \textit{privacy and security} issues. The last category they identified is \textit{user interface}, which includes an explanation of interface elements, especially if these have been changed. The survey data collected by Droste et al.~\cite{droste2024explanations} also served as the data basis for this work (as detailed in Section~\ref{sec:survey}).

Research into the user experience of explainable systems has shown that explanations can have a positive influence on aspects such as understandability~\cite{id955_dominguez2020,id1313_muhammad2016}, satisfaction~\cite{id731_bellini2018,id770_tran2019}, and trust~\cite{id731_bellini2018,id708_wang2018}. However, explanations may also have negative effects on user experience~\cite{chazette2020explainability,deters2024UXandExplainability,nunes2017systematic}. For example, they can negatively influence trust, if too much or misleading information is given~\cite{kizilcec2016trustAndTransparency,sadeghi2024explanation}. In addition, the users' cognitive load can be unnecessarily increased by misplaced explanations~\cite{chazette2020explainability,nunes2017systematic}. Poorly implemented explanations can also have a negative impact on the overall user experience~\cite{deters2024UXandExplainability}. It is therefore essential to implement explanations carefully to avoid unwanted side effects. Among other things, the personalization of explanations is a key factor here. For instance, the users' prior knowledge should be taken into account, as well as personal preferences based on the user's personality~\cite{deters2024qualitymodel}. To this end, Ramos et al.~\cite{ramos2021modeling} created personas to support the development of explainable systems.

Oberste and Heinzl~\cite{oberste2023explainabilityWithPriorKnowledge} explored how knowledge-informed machine learning enhances user-centric explanations in healthcare. Through a review of such systems, they highlighted improvements in formal understanding, medical knowledge delivery, and explanation intuitiveness. However, they noted the need for further research to tailor explanations to users’ backgrounds and improve trust and acceptance among medical professionals.

Unterbusch et al.~\cite{unterbusch2023explanation} investigated users' explanation needs in software systems by analyzing 1,730 app reviews from eight apps to develop a taxonomy of explanation needs. They also tested automated detection methods, achieving a weighted F-score of 86\% in identifying explanation needs across 486 reviews from four additional apps.

\subsection{Mood}
\label{sec:sentiment}
Mood is commonly measured with scales, such as the one created by Bohner et al. \cite{bohner1991stimmungs}, a German adaptation of Underwood and Froming’s "Mood Survey" \cite{underwood1980mood}. This scale assesses general, enduring mood \cite{bohner1997stimmungsskala} and includes two subscales: one for assessing enduring \textit{sentiment} and another for evaluating emotional \textit{reactivity} \cite{bohner1997stimmungsskala}, with reactivity measuring the strength of mood fluctuations. Comprising 15 statements rated on a 7-point scale \cite{bohner1997stimmungsskala}, the scale reverses negative polarity items during evaluation to ensure accuracy \cite{bohner1997stimmungsskala}, where higher values reflect a more positive mood or stronger emotional reactivity \cite{bohner1997stimmungsskala}.

For assessing mood over a defined period of time, such as over the past week, the \textit{Positive and Negative Affect Schedule} (PANAS) \cite{watson1988development} or its German version by Breyer and Bluemke \cite{breyer2016panas-de} is used. The scale comprises two 10-item subscales, one measuring positive affect (PA) and the other measuring negative affect (NA). Respondents indicate the extent to which they have experienced a given emotion on a 5-point Likert scale. PA reflects feelings of enthusiasm and high energy, whereas NA represents distress and unpleasurable emotions. It is a widely used measure in psychological research to assess emotional states and their implications for mental health.

Emotion classification frameworks categorize higher-level emotions, to which lower-level emotions are assigned~\cite{shaver1987emotion}. Shaver et al.~\cite{shaver1987emotion} developed one such framework based on data compiled by psychology students.

\section{Study Design}
\label{sec:research}
To investigate the relationship between mood, demographic factors, and explanation needs, we performed correlation analyses on an existing dataset from a previously conducted survey~\cite{droste2024explanations}.

\subsection{Research Goal and Research Questions}

We strive to achieve the following goal, formulated according to the Goal-Definition-Template~\cite{wohlin2012experimentation}:\\

\setlength{\shadowsize}{2pt}
\noindent
\shadowbox{
\begin{minipage}[t]{0.95\columnwidth}
\textbf{Research Goal:} \textit{Analyze} the relationship between user mood, demographic factors, and explanation needs 
\textit{for the purpose of} understanding how mood profiles and demographic attributes influence the frequency and types of explanation needs 
\textit{with respect to} uncovering patterns that inform user-specific explanation requirements 
\textit{from the point of view of} a researcher 
\textit{in the context of} an analysis of existing survey data.
\end{minipage}
}\vspace{0pt}

We investigate the following research questions:

\begin{itemize}
\item \textbf{RQ1: How does mood relate to explanation need?} 
Examining this question allows us to explore whether users’ emotional states influence their need for explanations in software, which could inform the development of adaptive support systems that dynamically respond to users’ moods.
To capture mood in a structured manner, we focus on sentiment (overall mood disposition) and reactivity (mood fluctuation intensity), as both aspects could influence users’ cognitive processing and, consequently, their need for explanations.

\item \textbf{RQ2: How does demographics relate to explanation need?} 
This question investigates the role demographic factors play in shaping explanation needs, offering insights into how personalized user support can be designed based on such attributes.

\end{itemize}

\subsection{Survey}
\label{sec:survey}
This paper uses data from the survey conducted by Droste et al.~\cite{droste2024explanations}. While the published study focuses on demographic data and explainability needs, the original survey~\cite{droste_2024_suppl} also included additional questions assessing sentiment and reactivity based on Underwood and Froming~\cite{underwood1980mood} (cf.~\ref{sec:sentiment}). However, only the questions relevant to the original study were publicly reported. The participants of the survey were acquired from the authors' personal and professional networks, either by directly contacting them or via \textit{LinkedIn}. Furthermore, the survey was posted on the authors' university's online message board. The relevant demographic data is detailed in Section~\ref{sec:demography}.

\subsection{Data Pre-Processing}
\label{sec:analyseaufbau}

To ensure the data could be used effectively in the correlation analysis, specific excerpts were extracted. Of the 83 survey participants~\cite{droste2024explanations}, 66 provided complete responses for mood and reactivity and were included in the final dataset.

Based on this structured dataset and the hypotheses, correlation analyses were conducted to explore potential relationships among explanation needs, their respective categories, and various demographic factors.

\subsection{Mood Analysis}

In this study, participants' mood was assessed using two dimensions: \textit{sentiment} and \textit{reactivity}. The survey items for these dimensions were adapted from Underwood and Froming~\cite{underwood1980mood} and are detailed in Table~\ref{tab:mood-trait}. These items capture aspects such as general happiness, consistency of mood states, and perceived emotional stability. Sentiment reflects the general positive or negative disposition of participants, while reactivity measures how quickly and frequently mood states change. These two dimensions were chosen as they represent both the stability (sentiment) and variability (reactivity) of mood, allowing us to investigate whether either influences explanation needs differently.

A 7-point Likert scale was employed for both sentiment and reactivity, ranging from strongly disagree (1) to strongly agree (7). This choice is relevant because a 7-point Likert scale provides a nuanced range of responses, allowing participants to express their mood and reactivity more precisely. The granularity of this scale captures subtle variations, distinguishing between participants who are slightly positive and those who are very positive. Such detailed data enable more robust statistical analyses, enhancing the understanding of the relationship between mood and explanation needs.

The items for each dimension were carefully designed to comprehensively evaluate mood traits, ensuring a thorough measurement of participants' overall emotional disposition and stability.

\begin{table}[htbp]
\small
\caption{Mood Survey Items by Underwood and Froming~\cite{underwood1980mood}}
\label{tab:mood-trait}
\begin{tabularx}{\textwidth}{XX}
\toprule
\textbf{Items for Sentiment} & \textbf{Items for Reactivity} \\
\midrule
\makecell[tl]{\textbullet~I usually feel quite cheerful.\\\textbullet~I'm frequently ``down in the dumps''.} & \textbullet~I may change from happy to sad and back again several times in a single week. \\
\makecell[tl]{\textbullet~I generally look at the sunny side of life.\\\textbullet~I'm not often really elated.} & \textbullet~Compared to my friends, I'm less up and down in my mood states. \\
\textbullet~I usually feel as though I'm bubbling over with joy. & \textbullet~Sometimes my moods swing back and forth very rapidly. \\
\makecell[tl]{\textbullet~I consider myself a happy person.\\\textbullet~I am not as cheerful as most people.} & \textbullet~My moods are quite consistent; they almost never vary. \\
\textbullet~Compared to my friends, I think less positively about life in general. & \textbullet~I'm not as ``moody'' as most people I know. \\
\textbullet~My friends often seem to feel I am unhappy. & \textbullet~I'm a very changeable person. \\
\bottomrule
\end{tabularx}
\end{table}

\subsubsection{Variables}
Based on the survey, we derived variables used for the data analysis. An overview of the variables used in our hypotheses can be found in Table~\ref{tab:variables-overview}. These variables are directly derived from the survey data~\cite{droste2024explanations,droste_2024_suppl}. 

\begin{table*}[h]
    \centering
    \setlength{\tabcolsep}{5pt} % Adjusting the spacing between columns
    \caption{Overview of Variables (NFE = Need for Explanation).}
    \label{tab:variables-overview}
    \begin{tabularx}{\textwidth}{lp{1.8cm}llX}
        \toprule
        \textbf{Variable}   & \textbf{Name} & \textbf{Scale} & \textbf{Range} & \textbf{Description} \\ 
        \midrule
        $d\textsubscript{gen}$ & Gender identity & Nominal & \{1; 2; 3\} & Female, male, diverse \\ 
        $d\textsubscript{age}$    & Age & Metric & $n \in \mathbb{N\textsubscript{0}}$ & Participant's age in years \\ 
        $d\textsubscript{react}$ & Reactivity & Ordinal & 1..7 & Measures the intensity of mood fluctuations on a 7-point scale \\ 
        $d\textsubscript{senti}$ & Sentiment & Ordinal & 1..7 & General mood level or "enduring mood," rated on a 7-point scale \\ 
        $E\textsubscript{total}$ & Overall NFE & Metric & $\mathbb{N\textsubscript{0}}$ & Total count of explanation needs expressed per person \\ 
        $E\textsubscript{need,X}$ & \makecell[l]{NFE by\\category X} & Ordinal & 1..5 & NFE within specific categories (e.g., functionality, data privacy) \\ 
        \bottomrule
    \end{tabularx}
\end{table*}

\subsubsection{Hypotheses Testing}
\label{subsec:hypothesen}
We examined the relationship between the need for explanations and demographic factors, sentiment and reactivity according to the null hypotheses shown in Table~\ref{tab:hypothesen-uebersicht}. 

\begin{table*}[htb]
\centering
\caption{Overview of the null hypotheses}
\label{tab:hypothesen-uebersicht}
\begin{tabularx}{\textwidth}{lXl}
\toprule
\textbf{Hypo.} & \textbf{Description (\enquote{There is no relation...})} & \textbf{Variables} \\ 
\midrule
H1\textsubscript{0}   & ... between demographic factors and sentiments. & $d\textsubscript{age}$, $d\textsubscript{gen}$, $d\textsubscript{senti}$ \\ 
H1.1\textsubscript{0} & ... between age and sentiments. & $d\textsubscript{age}$, $d\textsubscript{senti}$ \\ 
H1.2\textsubscript{0} & ... between gender identity and sentiments. & $d\textsubscript{gen}$, $d\textsubscript{senti}$ \\ 
\hline
H2\textsubscript{0}   & ... between demographic factors and reactivity. & $d\textsubscript{age}$, $d\textsubscript{gen}$, $d\textsubscript{react}$ \\ 
H2.1\textsubscript{0} & ... between age and reactivity. & $d\textsubscript{age}$, $d\textsubscript{react}$ \\ 
H2.2\textsubscript{0} & ... between gender identity and reactivity. & $d\textsubscript{gen}$, $d\textsubscript{react}$ \\ 
\hline
H3\textsubscript{0}   & ... between demographic factors and explanation need. & $d\textsubscript{age}$, $d\textsubscript{gen}$, $E\textsubscript{total}$ \\ 
H3.1\textsubscript{0} & ... between age and the overall explanation need. & $d\textsubscript{age}$, $E\textsubscript{total}$ \\ 
H3.1.X\textsubscript{0} & ... between age and the explanation need category X. & $d\textsubscript{age}$, $E\textsubscript{need,X}$ \\ 
H3.2\textsubscript{0} & ... in the overall explanation need between males and females. & $d\textsubscript{gen}$, $E\textsubscript{total}$ \\ 
H3.2.X\textsubscript{0} & ... in the explanation need category X between males and females. & $d\textsubscript{gen}$, $E\textsubscript{need,X}$ \\ 
\hline
H4\textsubscript{0}   & ... between sentiment and explanation need. & $d\textsubscript{senti}$, $E\textsubscript{total}$ \\ 
H4.1\textsubscript{0} & ... between sentiment and the overall explanation need. & $d\textsubscript{senti}$, $E\textsubscript{total}$ \\ 
H4.2.X\textsubscript{0} & ... between sentiment and the explanation need category X. & $d\textsubscript{senti}$, $E\textsubscript{need,X}$ \\ 
\hline
H5\textsubscript{0}   & ... between reactivity and explanation need. & $d\textsubscript{react}$, $E\textsubscript{total}$ \\ 
H5.1\textsubscript{0} & ... between reactivity and the overall explanation need. & $d\textsubscript{react}$, $E\textsubscript{total}$ \\ 
H5.2.X\textsubscript{0} & ... between reactivity and the explanation need category X. & $d\textsubscript{react}$, $E\textsubscript{need,X}$ \\ 
\bottomrule
\end{tabularx}
\end{table*}

\subsubsection{Correlation Analysis}
\label{sec:korrelationsanalyse}
To test the null hypotheses and analyze the relationship between the need for explanations, demographic factors, sentiment, and reactivity, we conducted correlation analyses. Pearson correlation coefficient $r$~\cite{cohen2009pearson} was used for testing relationships between two continuous variables. As gender is a nominal variable, we cannot use Pearson's $r$ to calculate the correlations. Instead, we opted for the Mann-Whitney-U test to compare the two groups (divided by gender) against each other. 

We tested the correlations for statistical significance at a significance level of $\alpha = 0.05$. When necessary, we adjust the significance level using the Bonferroni correction~\cite{haynes2013bonferroni} for main hypotheses. For example, if a main hypothesis is divided into two null hypotheses, the corrected significance level is at $\alpha\textsubscript{corr} = \frac{0.05}{2} = 0.025$. Any subhypothesis with a p-value below this threshold led to the rejection of the main hypothesis.

\section{Results}
\label{sec:results}

\subsection{Population Demography}
\label{sec:demography}

The analysis includes 66 participants from the dataset of Droste et al.~\cite{droste2024explanations}, as only these participants completed the entire sentiment questionnaire. Participants were aged between 18 and 72, with an average age of 38 and a median age of 28. Among these, 25 participants identified as female, and 41 identified as male. %The population distribution is displayed in Table~\ref{tab:population}.

The age distribution by gender is as follows. Among female participants, 12 individuals are in the 18–30 age group, 2 in the 30–40 age group, 2 in the 40–50 age group, 3 in the 50–60 age group, and 6 in the 60+ age group. Among male participants, 24 individuals are in the 18–30 age group, 5 in the 30–40 age group, 3 in the 40–50 age group, 2 in the 50–60 age group, and 7 in the 60+ age group.

\subsection{Survey Answers}

\subsubsection{Sentiment \& Reactivity}
Participants provided answers concerning their sentiments on 7-point Likert scales. For our analysis, we calculated an average sentiment score over the whole questionnaire for each participant. Sentiment scores ranged from 2.9 to 6.7, with a mean value of 4.8.

\begin{figure}[htbp!]
    \centering
    \includegraphics[width=4cm]{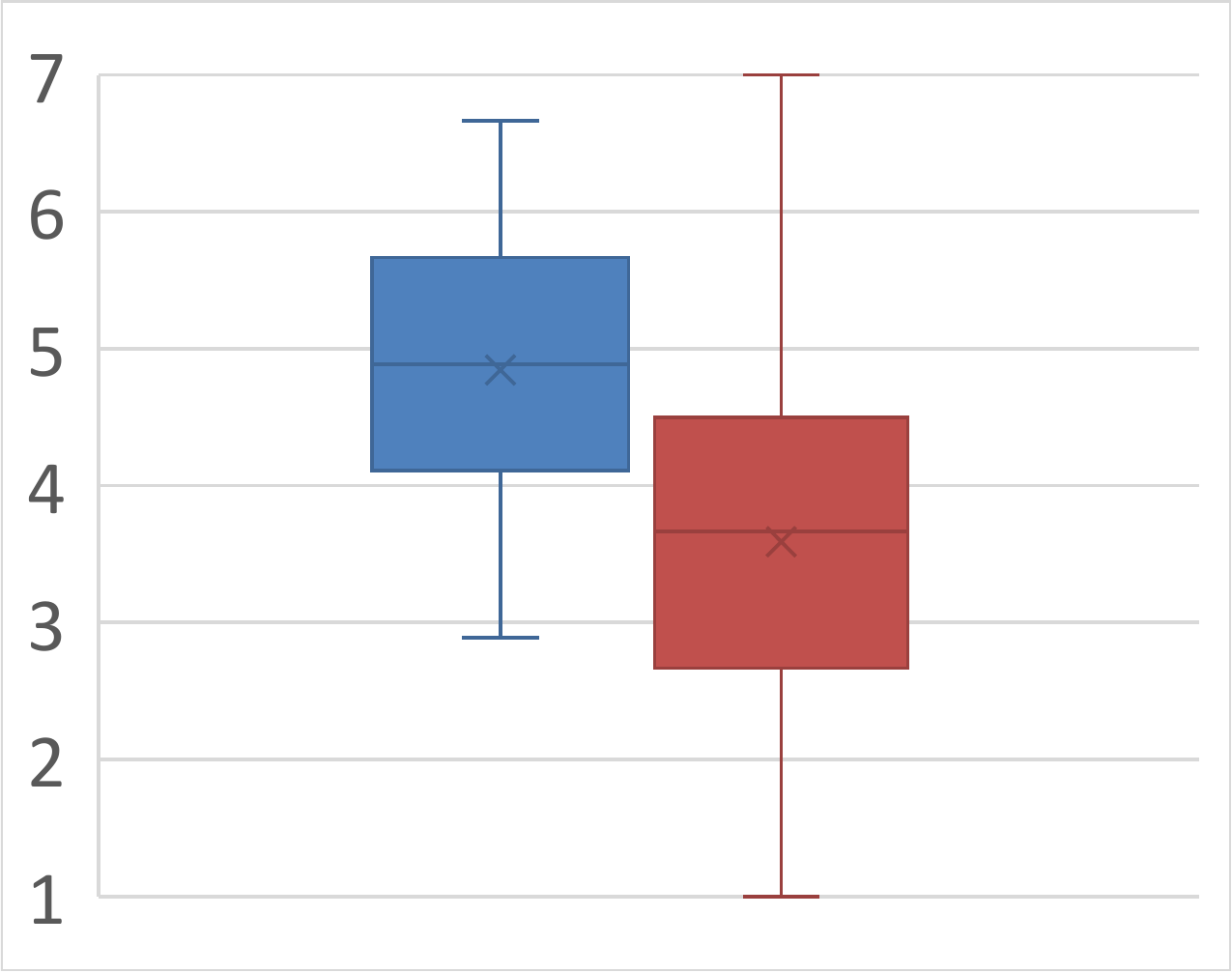}
    \caption{Distribution of sentiment (blue) and reactivity (red) among study participants.}
    \label{fig:sentimentandreactivity}
\end{figure}

Similarly, reactivity scores were derived from participant responses on a 7-point Likert scale, summarizing their self-reported intensity of mood changes. Reactivity scores ranged from 1 to 7, with a mean value of 3.6. The distribution of sentiment and reactivity among participants is visualized in Figure~\ref{fig:sentimentandreactivity}.

\subsubsection{Need for Explanations}
The sample of 66 participants reported an average of 4.6 explainability needs in their survey answers. The distribution of explainability needs across the sample is displayed in Table~\ref{tab:needs}. The distribution of needs in our sample aligns with the findings of Droste et al.~\cite{droste2024explanations}. Thus, we consider the sample is representative of the overall population in terms of expressed explanation needs.

\begin{table}[htb]
\centering
\caption{Number of needs}
\label{tab:needs}
\begin{tabularx}{\textwidth}{lXXXXX}
\toprule
\textbf{All}\;\; & \textbf{Interaction}\;\; & \textbf{Behavior}\;\; & \textbf{Domain}\;\; & \textbf{Security}\;\; & \textbf{Interface} \\ \midrule
240 & 118 (49.2\%) & 76 (31.7\%) & 21 (8.8\%) & 15 (6.3\%) & 10 (4.2\%) \\ \bottomrule
\end{tabularx}
\end{table}

\subsection{Correlation Analysis}
\label{sec:ergebnisse-der-korrelationsanalyse}
The results of the correlation analyses are summarized in Table~\ref{tab:hypothesen-analyse}. Regarding the relationship between sentimentand reactivity with explainability needs, we identified one statistically significant correlation. Reactivity and the need for user interface explanations showed a positive correlation at $r=0.33$, $p=0.007$, suggesting that participants with higher reactivity expressed greater needs for user interface explanations.

A significant correlation was observed between demographic factors and explainability needs. Specifically, older participants demonstrated fewer needs for user interface explanations ($r=-0.25$, $p=0.044$).

No significant correlations were identified between demographic data and sentiment or reactivity. Additionally, none of the main null hypotheses could be rejected after applying the Bonferroni correction.

\begin{table*}
\centering
\caption{Statistical analysis of the null hypotheses. For gender-based tests, Z-values from the Mann-Whitney-U test are reported.}
\label{tab:hypothesen-analyse}
\begin{tabularx}{\textwidth}{lXrr}
\toprule
\textbf{Hypothesis}   & \textbf{Variables of tested correlation} & \textbf{r-value / Z-value} & \textbf{p-value} \\
\midrule
H1\textsubscript{0}   & \textbf{Demographics \& sentiments.} & & \\
H1.1\textsubscript{0} & Age \& sentiments.              & 0.17 & .164 \\
H1.2\textsubscript{0} & Gender \& sentiments.              & 0.08 & .510 \\ \hline
H2\textsubscript{0}   & \textbf{Demographics \& reactivity.} & & \\
H2.1\textsubscript{0} & Age \& reactivity.              & -0.23 & .061 \\
H2.2\textsubscript{0} & Gender \& reactivity.              & 0.22 & .078 \\ \hline
H3\textsubscript{0}   & \textbf{Demographics \& expl. needs.} & & \\
H3.1\textsubscript{0} & Age \& overall expl. needs.              & -0.09 & .483 \\
H3.1.X\textsubscript{0} & Age \& interaction needs.              & -0.21 & .084 \\
H3.1.X\textsubscript{0} & Age \& behavior needs.              & -0.20 & .099 \\
H3.1.X\textsubscript{0} & Age \& domain needs.              & -0.06 & .632 \\
H3.1.X\textsubscript{0} & Age \& security needs.              & 0.03 & .799 \\
H3.1.X\textsubscript{0} & Age \& user interface needs.              & \textbf{-0.25} & \textbf{.044} \\
H3.2\textsubscript{0} & Gender \& overall expl. needs.        & Z = -1.86 & .063 \\
H3.2.X\textsubscript{0} & Gender \& interaction needs.              & Z = -1.26 & .208 \\
H3.2.X\textsubscript{0} & Gender \& behavior needs.              & Z = -0.67 & .503 \\
H3.2.X\textsubscript{0} & Gender \& domain needs.              & Z = -0.50 & .617 \\
H3.2.X\textsubscript{0} & Gender \& security needs.              & Z = -1.65 & .097 \\
H3.2.X\textsubscript{0} & Gender \& user interface needs.              & Z = -0.87 & .384 \\ \hline
H4\textsubscript{0}   & \textbf{Sentiment \& expl. needs.} & & \\
H4.1\textsubscript{0} & Sentiment \& overall expl. needs.              & -0.02 & .885 \\
H4.X\textsubscript{0} & Sentiment \& interaction needs.              & -0.13 & .299 \\
H4.X\textsubscript{0} & Sentiment \& behavior needs.              & 0.09 & .484 \\
H4.X\textsubscript{0} & Sentiment \& domain needs.              & -0.06 & .627 \\
H4.X\textsubscript{0} & Sentiment \& security needs.              & 0.09 & .469 \\
H4.X\textsubscript{0} & Sentiment \& user interface needs.              & -0.02 & .908 \\ \hline
H5\textsubscript{0}   & \textbf{Reactivity \& expl. needs.} & & \\
H5.1\textsubscript{0} & Reactivity \& overall expl. needs.              & 0.14 & .249 \\
H5.2.X\textsubscript{0} & Reactivity \& interaction needs.              & 0.15 & .225 \\
H5.2.X\textsubscript{0} & Reactivity \& behavior needs.              & 0.02 & .866 \\
H5.2.X\textsubscript{0} & Reactivity \& domain needs.              & 0.08 & .508 \\
H5.2.X\textsubscript{0} & Reactivity \& security needs.              & 0.09 & .483 \\
H5.2.X\textsubscript{0} & Reactivity \& user interface needs.              & \textbf{0.33} & \textbf{.007} \\
\bottomrule
\end{tabularx}
\end{table*}

\section{Discussion}
\label{sec:discussion}

In the following, we answer the research questions, present threats to validity, and interpret the results.

\subsection{Answering the Research Questions}
\label{sec:beantworten-der-forschungsfragen}

\textbf{RQ1: How does mood relate to explanation need?}
Our analysis revealed that mood, specifically emotional reactivity, has a significant relationship with explanation need, albeit only within specific contexts. We found a positive correlation between reactivity and the need for explanations related to user interfaces ($r=0.33$, $p=0.007$). This suggests that individuals with higher reactivity, who experience more intense emotional fluctuations, are more likely to require explanations in interface-related areas. However, no significant correlation was found between general sentiment and any category of explanation needs, indicating that while emotional reactivity influences the demand for explanations, a person's overall mood or sentiment does not have a direct impact.

\textbf{RQ2: How do demographics relate to explanation need?}
The results indicate that age, as a demographic factor, influences explanation needs. A significant correlation was observed between age and user interface explanation needs ($r=-0.25$, $p=0.044$), suggesting that younger users may require more support in understanding interface elements. This could be due to differing levels of familiarity with certain interfaces or varying expectations. No significant correlations were found between gender and explanation needs in this study.

\subsection{Interpretation}
\label{sec:interpretation}

The analysis reveals how demographic factors and mood impact explanation needs in software contexts. The following sections discuss key findings and their implications.

\textbf{Identified Correlations:}  
The analysis reveals minimal correlations between mood, demographic data, and explanation needs in software contexts. Among the tested hypotheses, only two significant correlations were found: between reactivity and user interface needs ($r=0.33$, $p=0.007$) and between age and user interface needs ($r=-0.25$, $p=0.044$). Participants with higher reactivity tended to express a greater need for user interface explanations, suggesting that emotional intensity may heighten users’ desire for clarity in navigating software interfaces. Similarly, younger participants demonstrated a higher demand for UI-related explanations, possibly reflecting differing levels of familiarity or expectations toward interface usability. This finding is somewhat unexpected, as younger users are typically perceived as more technologically adept. One possible explanation is that younger participants, having been exposed to a wider variety of digital interfaces, may have higher expectations regarding intuitive design and therefore perceive shortcomings more acutely, leading to an increased demand for explanations. Alternatively, older users might have developed more stable mental models for interacting with software over time, requiring less explicit guidance. 
These findings highlight the importance of considering both emotional and age-related factors when tailoring software explanations, particularly for younger, more reactive user groups.

\textbf{Importance of distinguishing sentiment from reactivity:}  
Our results underscore the need to differentiate between sentiment (a stable mood state) and reactivity (situational emotional intensity). While reactivity showed a significant correlation with explanation needs in UI contexts, sentiment did not correlate with any category of explanation needs. This suggests that situational emotional fluctuations may influence immediate explanation demands, whereas stable mood states have little to no impact. For researchers, this highlights the necessity of focusing on context-specific emotional responses rather than general mood when studying user behavior. For developers, designing adaptive systems that respond dynamically to real-time emotional feedback, such as increased reactivity, could enhance user satisfaction.

\textbf{Unexpected lack of broader correlations:}  
Contrary to expectations, no significant correlations were found between most demographic factors, mood dimensions, and explanation needs. From 28 null hypotheses, 26 could not be rejected, suggesting that explanation needs are not strongly influenced by these factors. These results are contrary to our initial assumption that subjective mood states or demographic characteristics like gender significantly shape explanation demands. It may indicate that explanation needs are more context-dependent or influenced by other user traits, such as cognitive style, familiarity with specific software, or intrinsic curiosity.

\textbf{Implications for research and industry:}  
The lack of significant correlations between mood, most demographic factors, and explanation needs suggests that software companies cannot reliably predict their users' explanation demands based on these attributes alone. This means that companies should actively gather explainability requirements directly from their users rather than relying on demographic or emotional profiles as proxies. While some factors, such as age or reactivity, showed limited correlations in specific contexts, these are insufficient for making generalized assumptions about users' needs.

For the industry, this means that a proactive approach is required: conducting user studies, collecting feedback through (online) surveys, interviews and workshops, and iteratively refining explainability features are still essential steps toward making software more transparent and accessible. Developers cannot assume that predefined user profiles will accurately capture the diversity of explanation needs within their user base.

\subsection{Future Work}
\label{sec:ausblick}

While this study provides valuable insights into the relationship between demographic and mood factors and users’ explanation needs, there are several avenues for future research that could deepen and expand these findings.

\textbf{Collecting user feedback through interviews or workshops:} 
To gain a more nuanced understanding of explanation needs, future studies could gather user feedback through structured interviews or workshops. Such qualitative methods would allow for in-depth exploration of specific explanation requirements, revealing subtleties that may not emerge from survey data alone. Conducting correlation analyses on this qualitative data could further illuminate how explanation needs vary with user characteristics, providing richer insights for more adaptive user support.

\textbf{Including additional personal factors:}
Expanding the scope of personal data to include variables such as life circumstances, psychological traits, and personality characteristics (e.g., openness, conscientiousness) could offer a more comprehensive view of what drives explanation needs. Incorporating these factors into analysis might help identify underlying personality traits or contextual factors that are more influential than mood alone. This could guide the development of highly personalized explanation systems that align with users' broader personal profiles.

\textbf{Exploring cultural influences through international studies:}
An international study would help determine whether cultural factors play a significant role in shaping explanation needs. By conducting similar research across diverse cultural contexts, future studies could assess whether certain explanation preferences are culturally specific or universal. This understanding would allow for the design of culturally adaptive explanation systems, enhancing usability and satisfaction in a global user base.

\textbf{Other analysis methods:}
Based on the correlation analysis, a regression analysis can provide more insights, as it allows for the simultaneous consideration of multiple variables in the analysis of their relationship with explanation need, whereas correlation analysis only compares variables individually~\cite{hosmer2013applied}.

\subsection{Threats to Validity}
\label{sec:validiteat}

In the following, we discuss threats to validity in our study, categorized according to the framework by Wohlin et al.~\cite{wohlin2012experimentation} as \textit{construct}, \textit{internal}, \textit{conclusion}, and \textit{external} validity.
Since this study is based on the work of Droste et al.~\cite{droste2024explanations}, the threats to validity outlined in their paper also apply here.

\textbf{Construct Validity.}
Our study’s construct validity is influenced by the specific mood data we collected, as we measured only one type of mood attribute. Additionally, since the data was collected through an online survey, there is a risk that participants may not have fully understood all (mood-related) questions. However, explanatory text and examples were provided to clarify the mood-related questions, which likely reduced misunderstandings. As our data and methodology closely follow the study by Droste et al.~\cite{droste2024explanations}, the threats to construct validity outlined in that work similarly apply to our study.

Survey items adapted from Underwood and Froming~\cite{underwood1980mood} may be interpreted differently due to individual differences in emotional granularity, potentially affecting sentiment and reactivity accuracy. Despite this, the 7-point Likert scale allows for nuanced responses and remains a structured approach to capturing mood variations.

\textbf{Internal Validity.}
Our correlation analysis only considers pairwise comparisons of variables, which limits our ability to detect more complex interdependencies between multiple factors. This approach, while standard, restricts the scope of the findings and could overlook underlying interactions among variables that influence explanation needs. Future studies could use multivariate analyses to explore such potential relationships more comprehensively. Furthermore, this work focused its demographic analysis on two factors: age and gender identity. Other demographic factors such as educational level or technical competence were not analyzed in this paper, even though they might yield interesting relationships.

Likert scales may introduce response biases in self-reported mood assessments. Some questions require subjective comparisons (e.g., "I am not as cheerful as most people"), making objective evaluation challenging. Additionally, Likert scales may oversimplify mood by relying on broad emotional categories, potentially affecting the accuracy of sentiment and reactivity measurements.

\textbf{Conclusion Validity.}
To minimize the likelihood of a Type I error (false positive), we applied the Bonferroni correction to our statistical tests. While this correction reduces the risk of incorrectly identifying significant relationships, it increases the chance of a Type II error (false negative), which may cause some true relationships to remain undetected. Moreover, we utilized Pearson and Mann-Whitney-U correlation methods to analyze the relationships between variables. Although these methods are established and appropriate for our variable types, they may not capture non-linear or more complex relationships. 

\textbf{External Validity.}
As our analysis is based on data from the study conducted by Droste et al.~\cite{droste2024explanations}, the external validity threats identified in their research similarly impact our study. Since our study’s findings are rooted in this specific dataset, they may not generalize to other settings or populations without further validation. Additionally, the study’s reliance on online survey responses may affect the generalizability of the results, as responses could vary in other survey environments or with a broader participant base.

\section{Conclusion}
\label{sec:conclusion}
In this work, we examined the relationship between users' need for explanations in software and factors such as mood and demographic data. Using data from a prior survey with 84 participants~\cite{droste2024explanations}, we analyzed responses from 66 participants who provided complete mood and demographic data. Pearson correlation coefficients and Mann-Whitney-U tests were applied to explore potential associations between these factors and specific types of explanation needs. Our results indicate a limited relationship between mood, demographic factors, and explanation needs. General mood (sentiment) did not show significant correlations with any category of explanation needs, challenging the assumption that stable mood states influence users’ desire for explanations. However, emotional reactivity was positively correlated with user interface explanation needs, suggesting that situational emotional intensity may heighten users’ demand for guidance in navigating user interfaces. Demographic data revealed only one significant correlation: age was negatively associated with user interface explanation needs. Younger users expressed a greater demand for UI-related explanations, potentially reflecting differing expectations or familiarity with software interfaces. No significant associations were observed between gender and explanation needs, suggesting that gender may not play as substantial a role in shaping explanation demands. The limited number of significant correlations emphasizes that explanation needs are highly subjective and cannot be reliably predicted based on mood or demographic data alone. This underscores the necessity for software companies to actively gather user-specific explainability requirements through direct engagement, such as surveys or workshops. Relying solely on inferred user profiles risks overlooking the diverse and context-dependent nature of explanation demands. Future work should explore additional personal traits, such as cognitive styles and personality characteristics, as well as cross-cultural influences, to provide a more comprehensive understanding of factors shaping explanation needs. Furthermore, we plan to validate our findings in industrial settings, by conducting workshops and focus groups with software practitioners. Such research will further inform the development of adaptive, user-centered systems that respond effectively to diverse user demands.

\section*{Acknowledgment}
This work was funded by the Deutsche Forschungsgemeinschaft (DFG, German Research Foundation) under Grant No.: 470146331, project softXplain (2022-2025).
\subsubsection*{Disclosure of Interests}
{\fontsize{9pt}{11pt}\selectfont The authors have no competing interests to declare that are relevant to the content of this article.}

\bibliographystyle{splncs04}
\bibliography{references.bib}

\begin{thebibliography}{10}
\providecommand{\url}[1]{\texttt{#1}}
\providecommand{\urlprefix}{URL }
\providecommand{\doi}[1]{https://doi.org/#1}

\bibitem{Andrade2019usercomplex}
Andrade, H., Lwakatare, L.E., Crnkovic, I., Bosch, J.: Software challenges in
  heterogeneous computing: A multiple case study in industry. In: SEAA (2019)

\bibitem{Antinyan2020complex}
Antinyan, V.: Revealing the complexity of automotive software. In: ESEC/FSE'20.
  Association for Computing Machinery (2020)

\bibitem{id731_bellini2018}
Bellini, V., Schiavone, A., Di~Noia, T., Ragone, A., Di~Sciascio, E.:
  Knowledge-aware autoencoders for explainable recommender systems. In: DLRS'18

\bibitem{bohner1997stimmungsskala}
Bohner, G., Schwarz, N.: Stimmungsskala. Zusammenstellung
  sozialwissenschaftlicher Items und Skalen (ZIS)  (1997)

\bibitem{bohner1991stimmungs}
Bohner, G., Hormuth, S.E., Schwarz, N.: Die stimmungs-skala: Vorstellung und
  validierung einer deutschen version des „mood survey “. Diagnostica
  \textbf{37}(2) (1991)

\bibitem{breyer2016panas-de}
Breyer, B., Bluemke, M.: Deutsche version der positive and negative affect
  schedule panas (gesis panel). Zusammenstellung sozialwissenschaftlicher Items
  und Skalen (ZIS)  (2016)

\bibitem{brunotte2023privacy}
Brunotte, W., Specht, A., Chazette, L., Schneider, K.: Privacy explanations--a
  means to end-user trust. JSS  \textbf{195} (2023)

\bibitem{chazette2021exploring}
Chazette, L., Brunotte, W., Speith, T.: Exploring explainability: a definition,
  a model, and a knowledge catalogue. In: RE. IEEE (2021)

\bibitem{chazette2022framework}
Chazette, L., Klös, V., Herzog, F., Schneider, K.: Requirements on
  explanations: A quality framework for explainability. In: RE (2022)

\bibitem{chazette2020explainability}
Chazette, L., Schneider, K.: Explainability as a non-functional requirement:
  challenges and recommendations. REJ  \textbf{25}(4) (2020)

\bibitem{cohen2009pearson}
Cohen, I., Huang, Y., Chen, J., Benesty, J., Benesty, J., Chen, J., Huang, Y.,
  Cohen, I.: Pearson correlation coefficient. Noise reduction in speech
  processing  (2009)

\bibitem{das2020XAISurvey}
Das, A., Rad, P.: Opportunities and challenges in explainable artificial
  intelligence {(XAI):}{A} survey. CoRR  \textbf{abs/2006.11371} (2020)

\bibitem{deters2023ondemand}
Deters, H., Droste, J., Fechner, M., Kl{\"u}nder, J.: Explanations on demand-a
  technique for eliciting the actual need for explanations. In: REW. IEEE
  (2023)

\bibitem{deters2024UXandExplainability}
Deters, H., Droste, J., Hess, A., Kl\"{o}s, V., Schneider, K., Speith, T.,
  Vogelsang, A.: The x factor: On the relationship between user experience and
  explainability. In: NordiCHI'24. Association for Computing Machinery

\bibitem{deters2024qualitymodel}
Deters, H., Droste, J., Obaidi, M., Schneider, K.: How explainable is your
  system? towards a quality model for explainability. In: REFSQ'24

\bibitem{id955_dominguez2020}
Dominguez, V., Donoso-Guzm\'{a}n, I., Messina, P., Parra, D.: Algorithmic and
  hci aspects for explaining recommendations of artistic images. ACM Trans.
  Interact. Intell. Syst.  \textbf{10}(4) (nov 2020)

\bibitem{droste2024explanations}
Droste, J., Deters, H., Obaidi, M., Schneider, K.: Explanations in everyday
  software systems: Towards a taxonomy for explainability needs. In: RE'24

\bibitem{droste_2024_suppl}
Droste, J., Deters, H., Obaidi, M., Schneider, K.: {Supplementary Material -
  "Explanations in Everyday Software Systems: Towards a Taxonomy for
  Explainability Needs" (RE'24)} (Mar 2024)

\bibitem{droste2023designing}
Droste, J., Deters, H., Puglisi, J., Kl{\"u}nder, J.: Designing end-user
  personas for explainability requirements using mixed methods research. In:
  REW. IEEE (2023)

\bibitem{Gabbas2021usercomplex}
Gabbas, M., Ryu, Y., Park, J., Kim, K.: Understanding challenges of designing
  for complex users by adapting the existing framework. In: Advances in
  Industrial Design. Springer International Publishing (2021)

\bibitem{haynes2013bonferroni}
Haynes, W.: Bonferroni Correction. Springer, New York, NY, USA (2013)

\bibitem{hosmer2013applied}
Hosmer, D.W., Lemeshow, S., Sturdivant, R.X.: Applied Logistic Regression. John
  Wiley \& Sons, Inc., Hoboken, NJ (2013)

\bibitem{Kim2023whatwhenexplain}
Kim, G., Yeo, D., Jo, T., Rus, D., Kim, S.: What and when to explain? on-road
  evaluation of explanations in highly automated vehicles. Proc. ACM Interact.
  Mob. Wearable Ubiquitous Technol.  \textbf{7}(3) (sep 2023)

\bibitem{kizilcec2016trustAndTransparency}
Kizilcec, R.: How much information? effects of transparency on trust in an
  algorithmic interface. In: CHI'16. Association for Computing Machinery

\bibitem{kortum2018impact}
Kortum, P., Oswald, F.L.: The impact of personality on the subjective
  assessment of usability. International Journal of Human--Computer Interaction
   \textbf{34}(2),  177--186 (2018)

\bibitem{kohl2019explainability}
Köhl, M.A., Baum, K., Langer, M., Oster, D., Speith, T., Bohlender, D.:
  Explainability as a non-functional requirement. In: RE'19

\bibitem{levy2021understanding}
Levy, O., Feitelson, D.: Understanding large-scale software systems –
  structure and flows. Empirical Software Engineering  \textbf{26}(1) (2021)

\bibitem{Mens2012complex}
Mens, T.: On the complexity of software systems. Computer  \textbf{45}(08) (aug
  2012)

\bibitem{id1313_muhammad2016}
Muhammad, K.I., Lawlor, A., Smyth, B.: A live-user study of opinionated
  explanations for recommender systems. In: IUI'16. Association for Computing
  Machinery

\bibitem{nunes2017systematic}
Nunes, I., Jannach, D.: A systematic review and taxonomy of explanations in
  decision support and recommender systems. User Modeling and User-Adapted
  Interaction  \textbf{27} (2017)

\bibitem{obaidi2025appKonwledge}
Obaidi, M., Fischbach, J., Herrmann, M., Deters, H., Droste, J., Kl{\"u}nder,
  J., Schneider, K.: How does users' app knowledge influence the preferred
  level of detail and format of software explanations? In: REFSQ'25 (2025)

\bibitem{obaidi2025automatingexplanationneedmanagement}
Obaidi, M., Voß, N., Droste, J., Deters, H., Herrmann, M., Fischbach, J.,
  Schneider, K.: Automating explanation need management in app reviews: A case
  study from the navigation app industry. ICSE-SEIP'25 (2025)

\bibitem{oberste2023explainabilityWithPriorKnowledge}
Oberste, L., Heinzl, A.: User-centric explainability in healthcare: A
  knowledge-level perspective of informed machine learning. IEEE Transactions
  on Artificial Intelligence  \textbf{4}(4) (2023)

\bibitem{ramos2021modeling}
Ramos, H., Fonseca, M., Ponciano, L.: Modeling and evaluating personas with
  software explainability requirements. In: HCI-COLLAB'21. Springer

\bibitem{sadeghi2024explanation}
Sadeghi, M., P\"{o}ttgen, D., Ebel, P., Vogelsang, A.: Explaining the
  unexplainable: The impact of misleading explanations on trust in unreliable
  predictions for hardly assessable tasks. In: UMAP '24. Association for
  Computing Machinery

\bibitem{shaver1987emotion}
Shaver, P.R., Schwartz, J.C., Kirson, D., O'Connor, C.: Emotion knowledge:
  further exploration of a prototype approach. Journal of personality and
  social psychology  \textbf{52 6} (1987)

\bibitem{id770_tran2019}
Tran, T.N.T., Atas, M., Felfernig, A., Le, V.M., Samer, R., Stettinger, M.:
  Towards social choice-based explanations in group recommender systems. In:
  UMAP'19

\bibitem{underwood1980mood}
Underwood, B., Froming, W.J.: The mood survey: A personality measure of happy
  and sad moods. Journal of Personality Assessment  \textbf{44}(4) (1980)

\bibitem{unterbusch2023explanation}
Unterbusch, M., Sadeghi, M., Fischbach, J., Obaidi, M., Vogelsang, A.:
  Explanation needs in app reviews: Taxonomy and automated detection. In: REW.
  IEEE (2023)

\bibitem{id708_wang2018}
Wang, X., Chen, Y., Yang, J., Wu, L., Wu, Z., Xie, X.: A reinforcement learning
  framework for explainable recommendation. In: ICDM (2018)

\bibitem{watson1988development}
Watson, D., Clark, L.A., Tellegen, A.: Development and validation of brief
  measures of positive and negative affect: the panas scales. Journal of
  personality and social psychology  \textbf{54}(6) (1988)

\bibitem{wohlin2012experimentation}
Wohlin, C., Runeson, P., Höst, M., Ohlsson, M.C., Regnell, B., Wesslén, A.:
  Experimentation in software engineering. Springer (2012)

\bibitem{xu2019roleOfUserMood}
Xu, L., Zhou, X., Gadiraju, U.: Revealing the role of user moods in struggling
  search tasks. In: SIGIR'19. Association for Computing Machinery

\end{thebibliography}

\end{document}